\documentstyle[12pt,epsf]{article}
\def\gsim{{~\raise.15em\hbox{$>$}\kern-.85em \lower.35em\hbox{$\sim$}~}}
\def\lsim{{~\raise.15em\hbox{$<$}\kern-.85em \lower.35em\hbox{$\sim$}~}}
\begin{document}
\begin{titlepage}
\vfill

\hskip 3.5in ISU-HET-97-2

\hskip 3.5in April, 1997

\hskip 3.5in Revised-August, 1997

\vskip 1.2in
\begin{center}
{\large \bf $CP$ Violation in $B$ Decays \\
from Anomalous $tbW$ Interactions}\\

\vspace{1 in}
{\bf A. Abd El-Hady and G.~Valencia}\\
{\it           Department of Physics and Astronomy,
               Iowa State University,
               Ames IA 50011}\\
\vspace{1 in}
\end{center}
\begin{abstract}

We calculate the effect of new $CP$ violating interactions 
parameterized by an anomalous $tbW$ coupling on $CP$-odd observables in 
$B$ decays. We find that couplings consistent with current bounds 
induce observable effects in some $CP$ asymmetries that will be 
measured in $B$-factories. The new effects are sufficiently large 
that they can actually test specific models that give rise to these 
$tbW$ interactions.

\end{abstract}
\end{titlepage} 
\clearpage

\section{Introduction}
 
The top quark is significantly heavier than the other five quarks. 
This has generated speculation that perhaps it plays a fundamental 
role in the breaking of electroweak symmetry \cite{ttbar,hill}. 
Models that incorporate this idea contain four-fermion operators 
that involve the third generation of quarks and perhaps exotic 
new fermions such as techniquarks.

At energy scales near the $W$ mass, this type of new theories 
gives rise to interactions between the $t$ and $b$ quarks and 
the electroweak gauge bosons $W$ and $Z$ that may deviate 
significantly from their standard model values. Such interactions 
are conveniently described by an effective Lagrangian \cite{pz}.

In this note, we study the effects of the simplest  $CP$-violating 
coupling in the effective Lagrangian to $CP$-odd observables in $B$
decays. The $CP$ conserving indirect effects of this coupling have 
been studied before in the literature \cite{raref,rare}. 
There are also studies of direct measurements of 
the $tbW$ coupling in future Tevatron experiments \cite{scott}. 
$CP$ violating interactions beyond the standard model have also been
studied in detail for  $B$ decays \cite{newcp}. However, the specific 
scenario that we discuss here has not been studied previously. 

\section{High Energy Effective Lagrangian}

We assume that whatever is responsible for generating
the non-standard model  top quark couplings occurs at a high energy 
scale $\Lambda$, perhaps a few~$TeV$. We also assume that this physics 
is responsible for the breaking of electroweak symmetry and that there 
is no light Higgs boson. Therefore, we use  a 
non-linear effective Lagrangian to describe the physics at the 
$W$ scale. Furthermore, in accordance with the prejudice that it is 
only the top-quark that plays a role in the new physics, we consider 
only the couplings of top and bottom quarks to $W$ and $Z$ gauge 
bosons.
  
To write the effective Lagrangian that 
describes the interactions of fermions to the electroweak gauge 
bosons we follow the formalism of Peccei and Zhang \cite{pz}. 
We consider only the lowest order couplings that can violate $CP$, 
and we do not include dipole moment type couplings of the top to the 
$Z$ since these have been studied before in the literature \cite{edmrev}. 
In unitary gauge we have:
\begin{equation}
{\cal L}_{\rm eff} = {g\over \sqrt{2}} V_{tb}
\biggl[(1+\kappa_L e^{i\phi_L}){\overline t}_L \gamma^\mu b_L+
\kappa_R e^{i\phi_R} {\overline t}_R \gamma^\mu b_R
\biggr] W_\mu^+ + h.c.
\label{ferml}
\end{equation}
The new interaction effects are parameterized by the real 
couplings $\kappa_L$ and $\kappa_R$ and the new phases $\phi_{L,R}$. 
These phases contain 
the information on $CP$ violation that may exist in the new theory. 
For simplicity we assume that 
the form Eq.~\ref{ferml} occurs only in the $tbW$ coupling ignoring 
any possible effects on $tsW$ and $tdW$ due to CKM mixing. The existing 
bounds on $\kappa_{L,R}$ depend on naturalness assumptions, so they 
are not rigorous \cite{rare}. To calibrate the sensitivity of the 
observables discussed in this paper, we will use the bounds 
$|\kappa_R| \lsim 0.01$ from $b \rightarrow s \gamma$ and \cite{raref},   
$|\kappa_L|\lsim 0.2$; 
obtained by setting any other anomalous coupling to zero in the results 
of Ref.~\cite{rare}. There are no bounds at present on the phases 
$\phi_{L,R}$.

Eq.~\ref{ferml} contributes to observables in $B$ decays at one-loop
order. However, we will not include in our calculation any additional 
effective operators that may be needed in a complete effective field
theory to act as counter-terms at one-loop. Instead we 
resort to estimating the order of magnitude of the effects by keeping 
either the one-loop contribution from Eq.~\ref{ferml} when it is 
finite; or the leading non-analytic term when it is divergent
\cite{georgi}. Therefore, our results 
will depend on the naturalness assumption that contributions 
from different couplings do not cancel each other. This is similar in 
spirit to the bounds that are placed on new gauge boson self
interactions from LEP observables \cite{dv}.

\section{Low Energy Effective Interactions}

In this section we present the results of the one-loop order contributions
of Eq.~\ref{ferml} to a low energy effective Lagrangian appropriate for
the study of $B$ decays. Two types of terms are generated corresponding 
to effective $|\Delta b|=1,2$ transitions.

For the $|\Delta b|=2$ transition we compute the usual box diagrams 
but with the $tbW$ coupling modified as indicated in Eq.~\ref{ferml}.  
In this case our result is divergent because the new interaction 
explicitly violates GIM. We keep the leading non-analytic term:
\begin{eqnarray}
H_{eff}&=& G_F^2 M_t^2 {(V^*_{tD}V_{tb})^2 \over 2\pi^2}\biggl\{ 
\biggl( H(x_t) -\kappa_L e^{i\phi_L}
\log\bigl({\mu\over M_W}\bigr) \biggr) \;
\overline{D}\gamma_\alpha P_Lb\overline{D}\gamma^\alpha P_Lb 
\nonumber \\
&-& 2\kappa_R^2 e^{2i\phi_R}   
\log\bigl({\mu\over M_W}\bigr) \; 
\overline{D} P_R b\overline{D} P_R b \biggr\}
\label{box}
\end{eqnarray}
We use $D$ to denote either a strange or down quark and,  
for comparison, we have included the standard model result as the first 
term. With $x_t = m_t^2/m_W^2$ we have \cite{inami}:
\begin{equation}
H(x_t)={3\over 2}{x_t^2\log x_t \over (x_t-1)^3}+
{x_t^2-11x_t+4 \over 4 (x_t-1)^2}
\label{smbox}
\end{equation}

For the $|\Delta b| =1$ transition we compute the standard gluonic  
penguin but use the complete interaction of Eq.~\ref{ferml} for the 
$tbW$ coupling. After we include the 
factors corresponding to wave function renormalization for the 
external fermions, we obtain the finite result:
\begin{eqnarray}
H_{eff}&=&{G_F \over \sqrt{2}}V^*_{tD}V_{tb}\biggl\{
{\alpha_S\over 4\pi}\biggl[F_L(x_t)+
\biggl(F_L(x_t)+{1\over 9}\biggr)\kappa_L e^{i\phi_L}\biggr] 
\; \overline{D}\lambda^a\gamma_\alpha
P_Lb\sum_q\overline{q}\lambda^a\gamma^\alpha q \nonumber \\
&+& {g_S\over 8 \pi^2}F_R(x_t)
\kappa_R e^{i\phi_R}\;m_t \; \overline{D}\sigma^{\mu\nu}\lambda^a P_R b 
G^a_{\mu\nu}\biggr\}\;+\;h.~c.
\label{loweff}
\end{eqnarray}
where,again, $D$ stands for either a strange or down quark, and the form
factors are:
\begin{eqnarray}
F_L(x_t) &=& -\log(x_t)
{(9x_t^2-16x_t+4) \over 6 (1-x_t)^4}  + 
x_t{(18-11x_t-x_t^2)\over 12(1-x_t)^3}\nonumber \\
F_R(x_t) &=& 3\biggl[\log(x_t){x_t\over (1-x_t)^3}+{(1+x_t)\over 2
(1-x_t)^2}+{1\over 6}\biggr]
\end{eqnarray}
The first term, independent of $\kappa_L$, corresponds to the standard 
model result and, not surprisingly, has the same form factor as the  
new left-handed coupling. 
The additional constant term,  $1/9$, is present because the new interaction 
does not have a GIM mechanism. For the same reason there is the term  
$1/6$ in $F_R$. It is interesting to note the potential for very large 
effects from the $\kappa_R$ term, where there is an enhancement factor of 
$m_t/m_b \sim 35$ compared to the corresponding operator in the 
standard model. This large factor can compensate for the smallness of 
$\kappa_R$ in the same way as it does in $b\rightarrow s \gamma$ 
transitions \cite{raref}.

\section{$CP$ Violation in neutral $B$ decays}

With the results of the previous section it is straightforward to 
estimate the effects of the new phases in the $CP$ asymmetries that will 
be studied in the $B$-factories. The general analysis of these $CP$ 
asymmetries has been reviewed in Ref.~\cite{brev}. 

The simplest way to estimate the potential size of the effect due to the
new phases is to look at processes with mixing induced $CP$ violation 
that are dominated by a tree-level amplitude. In this case 
the quantity of interest is (in the standard notation of \cite{brev}):
\begin{equation}
\biggl({q\over p}\biggr)_{B_D} = 
{V_{tb}^*V_{tD}\over V_{tb}V_{tD}^*} \biggl({F^* \over F}\biggr)^{1\over 2}
\end{equation}
where the first factor corresponds to the standard model value and 
the second factor is the modification due to the new phases:  
\begin{equation}
F=H(x_t)-\kappa_L e^{i\phi_L}\log\bigl({\mu\over M_W}\bigr)
-2\kappa_R^2 e^{2i\phi_R}\log\bigl({\mu\over M_W}\bigr)
{5\over 8}\biggl({M_B \over m_b}\biggr)^2
\end{equation}
The factor $5/8(M_B/m_b)^2$ for the right handed coupling takes into account
the difference in the hadronic matrix elements using factorization and
vacuum insertion. 
Numerically, we can use $m_t = 175~GeV$, and $\mu=1~TeV$ to find:  
\begin{equation}
\biggl({F^*\over F}\biggr)^{1\over 2} \approx 1+i\bigl[5\kappa_L\sin\phi_L+8\kappa_R^2
\sin(2\phi_R)\bigr] \equiv 1+i\phi_{Box}
\label{numbox}
\end{equation}
Experimental constraints on $\kappa_R$ from $b\rightarrow  s \gamma$
\cite{raref} make its contribution much smaller than that of
$\kappa_L$ in Eq.~\ref{numbox} so we will drop it. 

Since we assumed that the only new interaction is of the form shown in 
Eq.~\ref{ferml}, there is no corresponding modification for mixing in
the $K$ or $D$ systems. 
Strictly speaking, $\phi_{Box}$, may have a different value for 
$B_d$ and $B_s$ because, in principle, the two amplitudes may have 
different counter-terms. In our approximation we are ignoring the  
counter-terms and keeping only the leading logarithm 
so we end up with the same $\phi_{Box}$ for $B_d$ and $B_s$.

This is sufficient to study decay modes with tree-level dominated 
amplitudes (or after penguin effects are disentangled with an isospin 
analysis \cite{heo}). For
example, the modes $B_d\rightarrow \Psi K_s$ and $B_d \rightarrow \pi^+
\pi^-$ could be used to measure the angles 
$\beta$ and $\alpha$ \cite{brev}.\footnote{Alternatively one could use 
the modes $B \rightarrow \pi \rho$ to measure $\alpha$ \cite{pirho}.} 
With our new phases they
would really measure the combinations:
\begin{eqnarray}
\lambda (B_d \rightarrow \Psi K_s) &=& e^{-i(2\beta-\phi_{Box})} \nonumber \\
\lambda (B_d \rightarrow \pi^+ \pi^-) 
&=& e^{i(2\alpha+\phi_{Box})} 
\label{asymm}
\end{eqnarray}
Assuming that the phases $\alpha$ and $\beta$ are known 
with small theoretical uncertainties, for example 
from better measurements of $V_{ub}$ and $K^+ \rightarrow \pi^+ \nu 
\overline{\nu}$, one can 
use the $B$-factory measurements of $\alpha$ and $\beta$ to look for 
physics beyond the standard model. The BaBar technical design report 
quotes achievable 
errors on the measurements of $\alpha$ and $\beta$ of $8.5\%$ and 
$5.9\%$ respectively for $30fb^{-1}$ \cite{tdr}. 

The mode $B_d \rightarrow \Psi K_s$ is free of hadronic uncertainties 
both in the standard model \cite{brev}, and in our model. For hadronic 
uncertainties to appear in this mode, we would have to enhance penguin 
amplitudes by factors of at least 20. This is impossible for values of 
$\kappa_L$ and $\kappa_R$ that make sense in the context 
that we are discussing. Therefore, in the scenario in which $\beta$ 
is known and the $B$-factory measures $(\beta-\phi_{Box}/2)$ 
with a $5.9\%$ accuracy we can place the bound:
\begin{equation}
\kappa_L\sin\phi_L \lsim 0.03
\label{betabound}
\end{equation}
This is a very significant constraint since current $CP$ conserving 
data allows $\kappa_L$ to be as large as $\kappa_L \sim 0.2$ \cite{rare}. 

The mode $B_d \rightarrow \pi^+ \pi^-$ has hadronic uncertainties that 
may be resolved experimentally by carrying out an isospin analysis \cite{heo}. 
Assuming that this isospin analysis is possible and that the standard model 
is correct, this mode will give a measurement of  $\alpha$. 
Our model for new physics does not introduce 
new $\Delta I =3/2$ transitions, so the same isospin 
analysis would isolate the 
phase $(\alpha + \phi_{Box}/2)$. Since the experimental accuracy in 
this mode is worse than that of the previous mode, it will not lead 
to better bounds on $\kappa_L \sin\phi_L$. However, in our model 
the deviations in $\alpha$ and $\beta$ are 
related as in Eq.~\ref{asymm} and, thus, 
the value of $(\alpha + \phi_{Box}/2)$ is a prediction that could be tested.

We should point out that a value of $\kappa_L$ as large as 
$0.2$ would change the $B-\overline{B}$ mixing amplitude with respect  
to its standard model value by about $50\%$. This is well within the current 
theoretical uncertainty in the standard model calculation 
due to the hadronic form factors $f_B^2B_B$ \cite{fb}.

Decays in which penguin amplitudes are dominant offer the possibility 
to place bounds on $\kappa_R\sin\phi_R$. For penguin 
dominated modes one would have (using the same values of $m_t$ and $\mu$
as before and in the notation of \cite{brev}): 
\begin{equation}
{\overline{A}\over A} \approx \biggl({\overline{A}\over A}\biggr)_{SM}
\biggl(1+ 
i\bigl[ 3\kappa_L \sin\phi_L + 90 \kappa_R \sin\phi_R\bigr]\biggr)
\label{newpen}
\end{equation}
The large numerical factor in front of $\kappa_R$ is due to the $m_t/m_b$  
enhancement discussed earlier. To actually calculate this number 
one would have to be able to compute the hadronic matrix elements 
of the two operators in Eq.~\ref{loweff} and this is impossible at 
present. The number $90$ follows from a simple dimensional analysis 
in which we compare the two operators, imagine the gluon splitting into a 
quark-anti quark pair and replace any quark or gluon momentum with  
a factor of $M_B$. In this case the $\kappa_L$ term is less important 
so we drop it.

The general analysis of $CP$ asymmetries in $B$ decays with 
new physics in the decay amplitudes has been carried out in 
Ref.~\cite{yuval}. For a case like ours, it is convenient to 
compare the asymmetry in $B_d \rightarrow \Psi K_s$ with that in 
$B_d \rightarrow \phi K_s$. In models that only have new $CP$ violating phases 
in the mixing, the asymmetries in these two modes measure the same 
phase $(\beta+\delta_{Md})$ in the notation of Ref.~\cite{yuval}, (in our 
case $\delta_{Md} = -\phi_{Box}/2\approx -2.5\kappa_L\sin\phi_L$). 
With additional new phases in 
the decay amplitudes, the phase measured in $B_d \rightarrow \Psi K_s$
remains the same\footnote{Unless penguins are enhanced by factors 
of 20 or more, but we argue this does not happen in our model.},   
whereas the one measured in $B_d \rightarrow \phi K_s$ 
(or any other $b \rightarrow s\overline{s}s$ mode) becomes 
$(\beta+\delta_{Md}+\delta\phi_A)$ (again in the notation of 
Ref.~\cite{yuval}). From Eq.~\ref{newpen} we can read off the 
value $\delta\phi_A \approx 3\kappa_L \sin\phi_L + 90 \kappa_R \sin\phi_R$. 
Assuming that the difference in the phases of $b\rightarrow c\overline{c}s$ 
and $b\rightarrow s\overline{s}s$ modes can be measured to $10\%$ one could 
place the bound: 
\begin{equation}
\kappa_R\sin\phi_R \lsim 0.001
\label{boundpen}
\end{equation}
Even with the stringent bounds (of order a few percent) that $b \rightarrow 
s \gamma$ places on $\kappa_R$ \cite{raref}, this term can produce 
measurable corrections to CP asymmetries in penguin dominated modes. 
The bound in Eq.~\ref{boundpen} is extremely good, 
and would place constraints on models 
like one of Appelquist and Wu \cite{appel} where $\kappa_R\sin\phi_R$ 
can be as large as 0.01. 
The characteristic pattern of asymmetries induced by our anomalous 
couplings is the same as that of models with an enhanced chromomagnetic 
dipole operator of Ref.~\cite{kagan} that are discussed in 
Ref.~\cite{yuval}.\footnote{
It is difficult to be quantitative because we do not know how 
to calculate the necessary matrix elements, and because we don't yet 
know what will be the ultimate experimental 
accuracy. If 
we take $\kappa_R \sim 0.01$, our new interactions respect the 
CLEO bounds for $b\rightarrow s \gamma$ and for certain values of 
$\phi_R$ they can produce direct $CP$ violating asymmetries of order 
$10\%$ in modes 
where the standard model does not induce a sizable asymmetry such as 
$B \rightarrow \phi K$. We thank A.~Kagan who computed these numbers 
for us using his factorization model for the hadronic matrix elements 
\cite{alex}.}

\section{Conclusions}

In conclusion we have found that $CP$ asymmetries in $B$ decays are 
in principle sensitive to new $CP$ violating phases in the $tbW$ 
interaction. 
This is theoretically interesting because models of
electroweak symmetry breaking in which the top-quark plays a special
role may give rise to this type of interactions. Appelquist and Wu 
estimate in a technicolor model that $\kappa_R\sin\phi_R$ could be as 
large as 0.01 \cite{appel}, which we have seen is large enough to be 
seen at the $B$-factory. We have also seen that the $B$-factory can place 
bounds on $\kappa_L\sin\phi_L$ of order a few 
percent and that these bounds are 
meaningful in light of the current bounds on $\kappa_L$. 
Our result is 
phenomenologically interesting because the new $CP$ violating 
effects could be large enough
to distinguish them from standard model $CP$ violation in future experiments
at a $B$-factory. Finally, it is important to point out that the $B$-factory 
effects that we have discussed are likely to be the only place where one 
can search for new $CP$ violating phases of the type in Eq.~1. This is 
because the top couples so dominantly to bottom, that in direct top 
production and decay the phases cancel out as they enter both the 
production and decay vertices.

\section*{Acknowledgements}

The work of G. V. was supported in part
by the DOE OJI program under contract number DE-FG02-92ER40730. 
The work of A. Abd El-Hady was supported in part by the DOE under 
contract number DE-FG02-87ER40371. This work was supported 
in part by a LAS Faculty Development Grant from Iowa State University. 
We thank G. Burdman, Xiao-Gang~He, M. Hosch and D. London for conversations.
We are particularly grateful to Yuval Grossman for a critical reading of 
the manuscript and for bringing Ref.~\cite{yuval} to our attention, 
and to A.~Kagan 
for estimating the size of asymmetries that our model could induce in 
$b\rightarrow s\overline{s} s$ modes.

\end{document}